\documentclass[pra,aps,superscriptaddress,twocolumn,showpacs]{revtex4}
\usepackage{graphicx,amsfonts,bm,amsmath}
\newcommand{\kk}{{\bm k}}
\newcommand{\rl}{\rangle\!\langle}

\newcommand{\ff}{{\cal F}}

\begin{document}

\author{A. Grodecka}
\email{anna.grodecka@uni-paderborn.de}
\affiliation{Computational Nanophotonics Group, Theoretical Physics,
University Paderborn, 33098 Paderborn, Germany}
\author{P. Machnikowski}
\affiliation{Institute of Physics, Wroc{\l}aw University of Technology,
50-370 Wroc{\l}aw, Poland}
\author{J. F{\"o}rstner}
\affiliation{Computational Nanophotonics Group, Theoretical Physics, 
University Paderborn, 33098 Paderborn, Germany}

\title{Indirect spin dephasing via charge state decoherence in optical control schemes in~quantum dots}

\begin{abstract}
We demonstrate that an optically driven spin of a carrier
in a quantum dot undergoes indirect dephasing
via conditional optically induced charge evolution even in the absence
of any direct interaction between the spin and its environment. A generic model for the indirect dephasing 
with a three-component system with spin, charge, and reservoir is proposed.
This indirect decoherence channel is studied for the optical spin manipulation 
in a quantum dot with a microscopic description of the charge-phonon interaction
taking into account its non-Markovian nature.
\end{abstract}


\maketitle

\section{Introduction}

The idea of using quantum dots (QDs) for quantum computer implementations
follows from the possibility of a clear selection of a two level system,
on which a qubit can be realized \cite{loss98}.
To this end, both charge and spin states of confined carriers are employed,
where the latter is preferable, since spin states are generally more resistant
to decoherence processes.
Moreover, it is possible to exploit the charge evolution dependent on spin
(via selection rules and Pauli exclusion principle)
in order to manipulate the spin by optical means
\cite{imamoglu99,pazy03a,calarco03,gauger08} on picosecond time scales, that is, 
much faster than previously proposed magnetic or electrical control.
Many spin control schemes in such hybrid systems
which use off-resonant interband excitations together with 
STIRAP processes \cite{troiani03}, adiabatic \cite{chen04} and fast \cite{economou06,economou07}
evolution within trapped states in $\Lambda$
or four-level \cite{emary07a} systems have been proposed.
These hybrid systems are considered now as the most promising candidates for QD-based quantum computers
since during the millisecond spin decoherence time \cite{khaetskii01} 
it is possible to perform about $10^9$ optical quantum gates.
Optical rotation of a single spin performed via picosecond laser pulses with
the optical Stark effect as the operative mechanism 
was recently experimentally demonstrated \cite{berezovsky08}.
This pioneering experiment showed
that fast optical spin control is feasible and the current task is to
thoroughly study the decoherence mechanisms that limit the fidelity of
the achieved quantum control. 
The fundamental question is whether the spin degrees of freedom 
are indeed affected by decoherence mechanisms to a smaller degree than the charge ones
and what constitutes their main dephasing channel.

In this paper, we show that the spin state
of a confined carrier can undergo dephasing even in the absence
of spin-reservoir coupling if the spin rotation is achieved by a
conditional evolution induced on the orbital
degrees of freedom, as is the case in an optical control scheme. Although
the dynamical details of this dephasing process depend on the specific
implementation, the fundamental idea of the indirect
dephasing can be understood with the help of a ``generic model''
of a three-component system: the carrier spin, its orbital state, and the reservoir.
We show that this additional decoherence channel occurs on comparable or even shorter
timescales than the spin precession and trion decay during the optical manipulation.
Thus, it may constitute the main source of imperfections of the optical spin rotations.
This shows that phonon-induced dephasing should be included in the analysis of optical 
spin control schemes even though the commonly studied decoherence mechanism related to 
the material dependent spin-orbit coupling leads to very small errors 
for short gates \cite{khaetskii01} and indeed can be neglected.

The paper is organized as follows. In Sec.~\ref{sec:indirect}, a generic model describing
the indirect spin dephasing is introduced. Next, in Sec.~\ref{sec:model}, 
we present the model for the specific optical spin control protocol in a single QD.
Section \ref{sec:error} describes decoherence processes resulting
from carrier-phonon coupling. Section \ref{sec:concl} concludes the paper with final remarks.

\section{Indirect dephasing}\label{sec:indirect}

The idea of optical spin rotation is based on a spin-dependent evolution of
the charge, which finally brings it to the original state, up to an
additional phase accumulated during the evolution. 
Let the initial state be 
$|\psi(t_{0})\rangle = (\alpha|0\rangle_{\mathrm s} + \beta |1\rangle_{\mathrm s}) \otimes |0\rangle_{\mathrm c}$, 
where the components refer to spin (s) and charge (c) states, respectively.
The ideal evolution then has the form:
\begin{equation*}
|\psi_{\mathrm{id}}(t)\rangle = \alpha |0\rangle_{\mathrm s} \otimes |0\rangle_{\mathrm c}
+ \beta |1\rangle_{\mathrm s} \otimes \left[\eta(t) |0\rangle_{\mathrm c} + \xi(t)|1\rangle_{\mathrm c}\right],
\end{equation*}
where, at the final time $t_{1}$, $\eta(t_{1})=e^{i\phi}$ and
$\xi(t_{1})=0$. Typically, the occupation of the excited charge state is
kept small, $|\xi(t)|\ll|\eta(t)|$. This evolution realizes a rotation
of the spin by an angle $\phi$ around the axis defined by the states
$|0\rangle_{\mathrm s},|1\rangle_{\mathrm s}$, 
which may be selected at will using selection
rules and appropriate pulse phases and polarizations.

While the interaction between the spin and the environment is very
weak, there is much stronger scattering of the reservoir quanta on the
charge excitation, inducing, in a static situation, the usual phase damping channel
on the charge subsystem. In the present case, when the charge state
performs a conditional loop in its Hilbert space, the transient
occupation of the excited charge state leads to the accumulated
scattering amplitude (in the leading order in $\xi$ and $\epsilon$),
\begin{equation*}
w = i \epsilon\int_{t_{0}}^{t_{1}}dt |\xi(t)|^{2},
\end{equation*}
where $|\epsilon|^{2}$ is proportional to the scattering rate and we assume that the
reservoir quanta are non-resonant with the transitions between the
charge states (otherwise, additional leakage out of the computational
subspace appears). The final state of the three-component system is therefore
\begin{eqnarray*}
|\psi_{\mathrm{ac}}(t_1)\rangle &=&
\alpha|0\rangle_{\mathrm s} \otimes |0\rangle_{\mathrm c} \otimes |0\rangle_{\mathrm e} \\
&& + e^{i\phi} \beta |1\rangle_{\mathrm s} \otimes |0\rangle_{\mathrm c}
\otimes (\sqrt{1-|w|^{2}}|0\rangle_{\mathrm e} + w|1\rangle_{\mathrm e}),
\end{eqnarray*}
where the last component (e) represents the environment states. 
Thus, the charge state separates but the spin state becomes
entangled with the environment. Tracing out the charge and environment
degrees of freedom one arrives at the operator sum representation for
the effect of the imperfect rotation on the spin state,
\begin{equation*}
\rho_{\mathrm{ac}} = \sum_{\mu=0}^{1}M_{\mu}\rho_{\mathrm{id}} M_{\mu}
\end{equation*}
with $M_{0}=|0\rangle_{\mathrm{ss}}\langle 0|+\sqrt{1-|w|^{2}}|1\rangle_{\mathrm{ss}}\langle 1|$, 
$M_{1}=|w||1\rangle_{\mathrm{ss}}\langle 1|$.
In this way, the coupling between the
orbital degrees of freedom and the reservoir has induced an indirect
phase damping channel on the spin qubit (in the gate-dependent basis
$|0\rangle_{\mathrm s},|1\rangle_{\mathrm s}$), 
analogous to the indirect measurement scheme \cite{breuer02}
with the spin, charge and environment playing the roles of the
quantum object, quantum probe and measurement device, respectively.

In the following, we study in detail the indirect dephasing process
for a specific optical spin control protocol \cite{economou07}, including the
microscopic description of the interaction between charges and their
phonon reservoir as well as the non-Markovian nature of the latter. 
We show that this dephasing process leads to considerable errors, much
larger than those induced by the spin-orbit coupling or hyperfine
interaction over the relatively short gate duration on the picosecond timescale.

\section{Model system}\label{sec:model}

The considered system consists of a single QD doped with one electron. A magnetic field is applied
in the $x$ direction (Voigt configuration) and generates Zeeman splittings $2\omega_e$ 
between the two electron spin states $|\bar x\rangle$ and $|x\rangle$
with fixed spin projection on the $x$ axis equal to $-1/2$ and $+1/2$, respectively.
Analog for the trion spin states $|T_{\bar x}\rangle$ and $|T_{x}\rangle$
with energy splitting $2\omega_h$.
These states are linear combinations of the electron 
($|\bar z\rangle$, $|z\rangle$) and trion ($|\bar T\rangle$, $|T\rangle$) spin states 
along the growth and optical axis $z$. Depending on the light
polarization, rotations about different axes are accomplished. 

\begin{figure}[tb] 
\unitlength 1mm
{\resizebox{50mm}{!}{\includegraphics{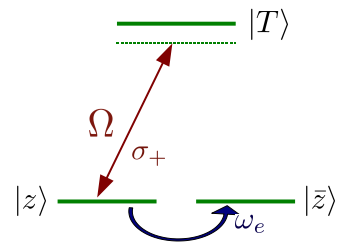}}}
\caption{\label{fig:lambda} $\Lambda$ system in a single quantum dot.}
\end{figure}

As shown in Ref.~\onlinecite{economou06}, a rotation about the $z$ axis is
performed with off-resonant circularly polarized light which,
according to selection rules, couples the two spin states to only one
trion state. Thus, we deal with an evolution of a three-level $\Lambda$ system 
(see Fig.~\ref{fig:lambda}).
The control Hamiltonian, including free carrier part and carrier-light
interaction, reads
\begin{eqnarray*}
H_{\mathrm C} & = & \omega_e (|z\rl \bar z| + |\bar z\rl z|) + \epsilon_T |T\rl T|\\
&& + \Omega_z(t) \left(e^{i\omega_z t} |z\rl T| + \mathrm{H.c.}\right),
\end{eqnarray*}
where the laser pulse couples only the one spin state $|z\rangle$ and a trion state $|T\rangle$,
whereas the orthogonal spin state $|\bar z\rangle$ is indirectly coupled via the magnetic field. 
After a passage of a $2\pi$ sech pulse, 
$\Omega_z(t) = \Omega_z \mathrm{sech}(\sigma_z t)$, the state acquires a phase, 
which, in consequence, leads to a spin rotation.
The angle of rotation, $\phi_{z} = 2 \arctan(\sigma_{z}/\Delta_z)$, 
is defined via the laser bandwidth $\sigma_{z}$
and detuning of the laser from the transition energy $\Delta_z = \epsilon_T - \omega_z$.
No population transfer to a trion state is possible for $\sigma_z = \Omega_z$.
The approximation made in this scheme requires that the spin is considered to be frozen 
during the pulse, i.e. $\sigma_z \gg \gamma$, where $\gamma = 2(\omega_e+\omega_h)$, 
which from the beginning imposes a limitation on driving conditions 
(short pulse durations especially for large Zeeman splittings).

The free phonon Hamiltonian has the form
$H_{\mathrm{ph}} = \sum_{\kk} \hbar \omega_\kk^{\phantom{\dagger}} \beta_\kk^\dagger \beta_\kk^{\phantom{\dagger}}$,
where $\kk$ is the phonon wave number and $\beta_\kk^\dagger$ ($\beta_\kk^{\phantom{\dagger}}$) 
is a phonon creation (annihilation) operator with corresponding frequencies $\omega_\kk$.
The Hamiltonian describing the interaction of the carriers with phonons reads
\begin{equation*}
H_{\mathrm{c-ph}} = \sum_{n,n'} |n \rl n'| \sum_{\kk} f_{nn'}(\kk) 
(\beta_\kk^{\phantom{\dagger}} + \beta_{-\kk}^{\dagger}),
\end{equation*}
where $f_{nn'}(\kk)$ are coupling elements and $n=z, \bar z, T$, and $\bar T$. 
The off-diagonal elements can be neglected
due to energetic reasons and low efficiency of direct phonon-assisted spin-flip processes.
Moreover, $f_{zz}(\kk)=f_{\bar z \bar z}(\kk)$ since the orbital wave functions are the same.
Before the pulse is switched on, the lattice is already in a new dressed equilibrium state \cite{machnikowski07a}
due to doping with one electron,
and the phonon modes can be redefined in terms of new operators $b_\kk = \beta_\kk + f_{zz}(\kk)/(\hbar \omega_\kk)$.
In the strong confinement regime, a trion state can be written in a product form of electron and hole states.
The resulting carrier-phonon Hamiltonian is 
\begin{equation*}
H_{\mathrm{c-ph}} = |T\rl T| \sum_{\kk} F_{TT} (\kk) \left(
b_\kk^{\phantom{\dagger}} + b_{-\kk}^{\dagger} \right)
\end{equation*}
with the following deformation potential coupling element 
between a trion and the phonon environment~\cite{grodecka07}
\begin{equation*}
F_{TT}(\kk) = f_{TT}(\kk) - f_{zz}(\kk) = \sqrt{\frac{\hbar k}{2\rho V c_l}}
(D_e - D_h) \ff (\kk).
\end{equation*}
Here, $\rho = 5360$~kg/m$^3$ is the crystal density, $V$ is the normalization volume of the phonon modes,
$c_l = 5150$~m/s is the longitudinal speed of sound, $\ff (\kk)$ is the form factor reflecting the geometrical
properties of the wave functions \cite{grodecka08}, and $D_e$ ($D_h$) is the deformation potential constant
for electrons (holes), where $D_e - D_h = 8$~eV. 
These parameters correspond to self-assembled InAs/GaAs
quantum dots with the electron and hole confinement in-plane equal to $4$~nm and in growth direction $1$~nm.

\begin{figure}[tb] 
\unitlength 1mm
{\resizebox{90mm}{!}{\includegraphics{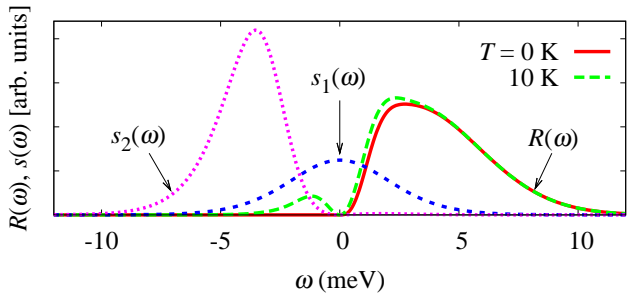}}}
\caption{\label{fig:spec} Phonon spectral density $R(\omega)$ at two temperatures 
and two spectral characteristics of the driving $s_1(\omega)$ and $s_2(\omega)$ 
for $\pi/2$ rotation about the $z$ axis with detuning and pulse bandwidth $\sigma_z = \Delta_z =  2.6$~meV.}
\end{figure}

\section{Phonon-induced decoherence}\label{sec:error}

To measure the quality of the operation on a qubit we use the 
\textit{error} of the quantum gate, $\delta = 1-F^2$,
defined as the loss of fidelity $F$. 
The error is a difference between the ideal final state (without decoherence)
and the actually achieved one including the coupling to environment.
Here, we consider the interaction with phonon environment,
however, the trion radiative coupling (carrier-photon interaction)
can be described in the same manner.
The effect of the interaction with the phonon reservoir 
is calculated via the second order Born expansion of the density matrix evolution equation 
(for details, see Ref.~\onlinecite{grodecka07}).
The interaction with light is included exactly and coupling to phonons 
is treated within a non-Markovian perturbation theory.
As a result, one can write the error of the quantum gate as an overlap between two spectral functions
reflecting the properties of the two above interactions,
\begin{equation*}
\delta = \int d\omega R(\omega) S(\omega).
\end{equation*}
Here, 
\begin{equation*}
R(\omega) = \frac{n_B+1}{\hbar^2} \sum_\kk |F_{TT}(\kk)|^2
[\delta(\omega - \omega_\kk) + \delta(\omega + \omega_\kk)]
\end{equation*}
is the phonon spectral density representing phonon emission ($\omega>0$)
and absorption ($\omega<0$, nonzero only at finite temperature) processes (see Fig.~\ref{fig:spec}).

\begin{figure}[tb] 
\unitlength 1mm
{\resizebox{80mm}{!}{\includegraphics{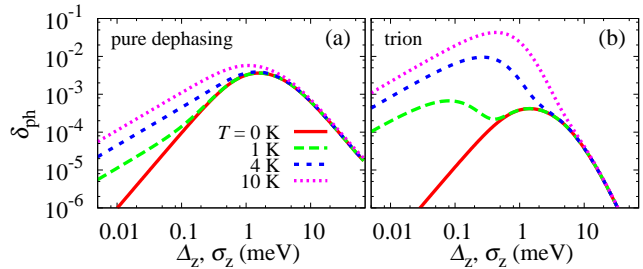}}}
\caption{\label{fig:pi0.5} Phonon-induced error contribution due to 
(a) pure dephasing and (b) phonon-assisted trion generation during the $\pi/2$ rotation about the $z$ axis.}
\end{figure}

The spectral characteristics of the driving, $S(\omega)$, has as many contributions as the 
dimension of the orthogonal complement of the initial state. In the case of $z$ rotation, there are
two contributions, $S(\omega) = s_1(\omega) + s_2(\omega)$ reflecting two phonon-induced decoherence channels. 
One represents pure dephasing mechanism and reads
\begin{equation*}
s_{1}(\omega) = \frac{1}{4} \sin^{2}\vartheta \left| \int dt \; e^{-i\omega t} \;
\left|-\frac{i}{c^*} \xi^{c^*} (1-\xi)^c \right|^2 \right|^{2},
\end{equation*}
where $c = (1+i\Delta_z/\sigma_z)/2$ and the time dependence is enclosed in 
$\xi(t) = [ \tanh(\sigma_{z}t) +1 ]/2$.
This function is always centered at $\omega=0$ (Fig.~\ref{fig:spec}) and its
width grows with growing pulse bandwidth and detuning.
It results from the fact, that the dynamical errors depend on the evolution speed,
i.e., for a given pulse duration only some phonon modes can follow the evolution
adiabatically whereas the others relax contributing to dephasing.
The same applies to the second spectral function
\begin{equation*}
s_2(\omega) = \cos^{2}\frac{\vartheta}{2} \left| \int dt \; e^{-i\omega t} \; 
\frac{(-i)}{c^*} \xi^{c^*} (1-\xi)^c \left( 1 - \frac{\xi}{c^*} \right) \right|^{2},
\end{equation*} 
but the center of this function is shifted to a negative frequency around detuning value $\omega \approx -\Delta_z$. 
This contribution represents real transition and constitutes a decoherence channel referred to as
the phonon-assisted trion generation.

The resulting phonon-induced errors during a $\pi/2$ rotation about the $z$ axis,
averaged over all initial spin states, are plotted 
in Fig.~\ref{fig:pi0.5} as functions of detuning and bandwidth 
(in this case, $\Delta_z = \sigma_z$) at four different temperatures $T$.

The first contribution to the error resulting from pure dephasing effects [Fig.~\ref{fig:pi0.5}(a)] 
initially grows with growing detuning and pulse bandwidth.
For small pulse bandwidths, the evolution is really slow and the relevant function $s_1(\omega)$
is extremely narrow covering only the diminishing part of the phonon density at $\omega\approx 0$.
Thus, the phonons are able to adiabatically follow the change of the charge distribution
and as a result the decoherence is reduced.
Unfortunately, the proposed schemes require usually bandwidths much larger than Zeeman splitting
and one cannot use the discussed bandwidth sector with small errors.
This error contribution reaches its maximum value for $\Delta_z = \sigma_z \approx 1.5$~meV
for all temperature values, where the pure dephasing effects are most efficient 
[$s_1(\omega)$ is broad and covers the whole spectrum of phonons].

The second error due to phonon-assisted transitions to the trion state
is plotted in Fig.~\ref{fig:pi0.5}(b).
In this case, the temperature dependence is stronger, since the spectral characteristics is
centered at the negative frequency part of the phonon spectral density, which is strongly temperature dependent.
Even for small bandwidths at a relatively low temperature $T=1$~K,
the error is larger than $10^{-4}$. At each temperature, the maximum error is reached for
the detuning corresponding to the maximal value of the phonon density. 
The error diminishes for large detunings ($>50$~meV) after the spectral characteristics reaches the phonon cut-off,
where the one-phonon processes are not efficient.

\begin{figure}[tb] 
\unitlength 1mm
{\resizebox{85mm}{!}{\includegraphics{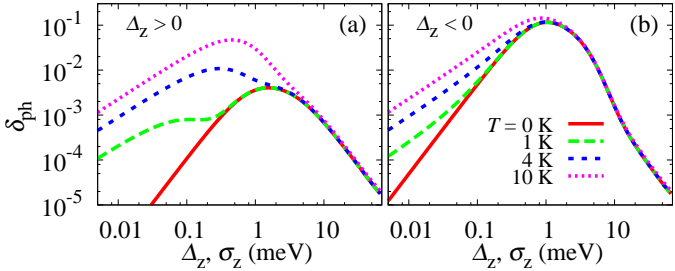}}}
\caption{\label{fig:pi0.5tot} Total phonon-induced error for 
(a) positive and (b) negative detuning during the $\pi/2$ rotation about the $z$ axis.}
\end{figure}

The total phonon-induced error during the $\pi/2$ rotation about growth direction $z$
is plotted in Fig.~\ref{fig:pi0.5tot}(a). 
To guarantee the coherent control and reach small errors, 
one needs either very small values of detunings and pulse bandwidth
or very large ones of a few tens of meV. 
Taking into account the bandwidth limitation for typical Zeeman splitting of $0.1$~meV,
the available parameters lead to large gate errors even at zero temperature.
The only way to obtain desired small errors is to use very large detunings and short pulse durations.
However, under such conditions, many other decoherence channels like
resonant and off-resonant transitions to higher states or
interaction with optical phonons are likely to appear. 
Moreover, this can lead to experimental difficulties, since large detunings require very strong pulses.

In order to perform a rotation about an arbitrary axis, rotations about two orthogonal axes are needed,
e.g. $z$ and $x$, and detunings above the energy gap may be needed. 
This leads to much larger phonon-induced errors, since emission processes become very important here.
The total phonon-induced error for negative detunings is plotted in Fig.~\ref{fig:pi0.5tot}(b).
Now, the spectral characteristics $s_2(\omega)$ responsible for phonon-assisted trion generation
is centered at positive frequencies, where the phonon spectral density 
has much larger values especially at low temperature.
In this case, the errors are up to two orders of magnitude larger 
in comparison with those for positive detunings. 
For experimentally reasonable values of detunings and pulse bandwidth, 
the error is always larger than $10^{-2}$ and has the maximal value of $\approx 10^{-1}$
for $\Delta_z = \sigma_z \approx 1$~meV.

\begin{figure}[t] 
\unitlength 1mm
{\resizebox{70mm}{!}{\includegraphics{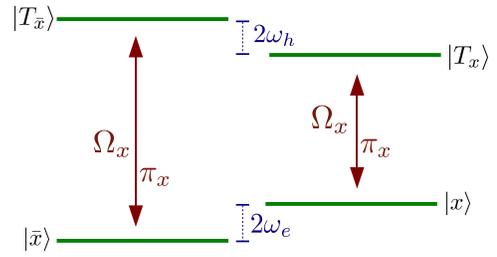}}}
\caption{\label{fig:4lev} $4-$level system in a single quantum dot.}
\end{figure}

\begin{figure}[b] 
\unitlength 1mm
{\resizebox{70mm}{!}{\includegraphics{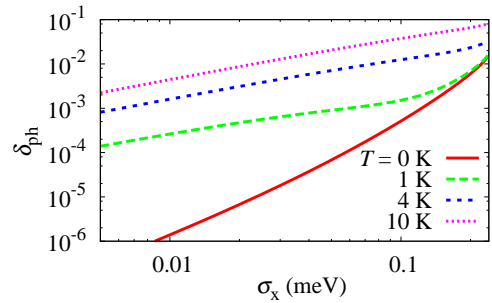}}}
\caption{\label{fig:xrot} Total phonon-induced error for the $\pi/2$ rotation about the $x$ axis.}
\end{figure}

The spin rotation about the $x$ axis is realized via linearly polarized $\pi_{x}$ pulse.
The relevant control Hamiltonian is:
\begin{eqnarray*}
H_{{\rm C}x} & = & \omega_e (|x\rl x| - |\bar x\rl \bar x|)
+ \epsilon_T^{(-)} |T_x\rl T_x| + \epsilon_T^{(+)} |T_{\bar x}\rl T_{\bar x}| \\ 
&& + \left[ \Omega_x(t)e^{-i\omega_x t} (|x\rl T_x| + |\bar x\rl T_{\bar x}|)
+ \mathrm{H.c.} \right],
\end{eqnarray*}
where $\epsilon_T^{(\pm)} = \epsilon_T \pm \omega_h$.
In this case, all four levels participate in the evolution (see Fig.~\ref{fig:4lev}).
In consequence, there are two paths for phonon-assisted trion generation
with two different detunings.
The resulting total phonon-induced gate error for the $\pi/2$ rotation about the $x$ axis
is shown in Fig.~\ref{fig:xrot}. One can see, that already at $T=1$~K, 
the error is always larger than $10^{-4}$ and grows with growing bandwidth.

Adding the individual errors, it is possible to estimate the error of an arbitrary spin rotation.
As we already discussed, even for a rotation about one of the axes,
it is impossible to find driving conditions leading to
errors smaller than $10^{-4}$, thus for the $y$-rotation the situation is even worse.
Moreover, the calculated errors for single-qubit gates provide an estimation
for the two-qubit spin gates employing, for instance, 
the electron hole exchange interaction in coupled QDs \cite{economou08}.

\section{Discussion and conclusions}\label{sec:concl}

It has been shown that even in the absence of direct spin-reservoir coupling,
the spin state of a confined carrier is exposed to indirect dephasing 
through the entangling optically induced charge evolution. We have proposed a model
for this indirect decoherence channel consisting of three components:
spin, charge, and environment. As an illustration, the optical spin manipulation
in a single doped quantum dot has been considered. 
It was shown, that optical driving of such a system
leads to a strong dynamical response of the lattice and to strong
indirect dynamical phonon-induced decoherence channels for the spin degrees of freedom.

Finally, we compare the considered optical spin control proposal 
with two previous schemes \cite{troiani03,chen04}.
All of them use single or double quantum dots doped with one additional electron
and the excitation of intermediate trion state.
One of them \cite{troiani03} makes use of stimulated Raman adiabatic passage (STIRAP)
and is implemented in a double quantum dot.
The main limitations of this proposal are slow adiabatic evolution requirement
and necessity of electron transfer between two QDs and the delocalized hole state.
The second one \cite{chen04} is implemented in a single QD so that the two latter constrains
are overcome. However, the evolution still has to be adiabatic.
The proposal considered in this paper prevails over all the limitations discussed above,
since the optical rotation is performed by means of fast laser pulses.
However, this leads to larger phonon-induced errors $\delta > 10^{-3}$ even at low temperature,
whereas the errors in the case of adiabatic evolution \cite{roszak05b,grodecka07}
are at least one order of magnitude smaller $\delta < 10^{-4}$.
On the other hand, the fast evolution leads to smaller errors resulting from 
carrier-photon interaction. The trion state is excited only for a short moment,
thus the probability of its radiative decay is low \cite{grodecka07}.
All in all, the fast optical spin rotation analyzed here possesses many advantages
in comparison with the other two proposals, however the dynamical phonon-induced
indirect spin dephasing is in this case much stronger.

The phonon-induced decoherence processes may in many cases constitute the dominant source of errors,
since they are much more efficient than those due to
spin-orbit mechanism assisted by phonons and up to two orders of magnitude larger than the errors resulting
from trion radiative decay \cite{grodecka07}. Moreover, these dynamical phonon-induced processes 
are most efficient exactly on the timescales for the proposed and demonstrated optical spin rotations.
Therefore, in order to overcome the phonon-induced indirect spin dephasing 
one should avoid such detunings and timescales. Another idea is to reduce the dephasing by means
of collective encoding of quantum information in QD arrays \cite{grodecka06,grodecka06b}.
Pulse optimization may also lead to error reduction \cite{axt05a,wenin06,hohenester04}.

\acknowledgments
A. G. and J. F. acknowledge support from the Emmy Noether Program of the DFG (Grant No. FO 637/1-1)
and the DFG Research Training Group GRK 1464, 
and thank the John von Neumann Institut f{\"u}r Computing (NIC) for computing time.

\bibliographystyle{prsty}
\bibliography{abbr,quantum}

\end{document}